\documentclass[11pt,a4paper]{article}

\usepackage[utf8]{inputenc}
\usepackage[T1]{fontenc}
\usepackage{lmodern}
\usepackage[margin=1in]{geometry}
\usepackage{amsmath}
\usepackage{amssymb}
\usepackage{booktabs}
\usepackage{graphicx}
\usepackage{xcolor}
\usepackage[numbers,sort&compress]{natbib}
\usepackage{caption}
\usepackage{enumitem}
\usepackage[hidelinks]{hyperref}

\graphicspath{{figures/}}

\newcommand{\fpr}{\mathrm{FPR}}
\newcommand{\Tnull}{T_{0}}
\title{\bfseries The Limits of Behavioral Early Warning for\\
Copy-Trading Blow-Ups\\[4pt]
\large A Pre-Registered, Full-Universe On-Chain Study (Hyperliquid)}

\author{Syuan Wei Peng (Sean Peng)\thanks{Mnemox AI. Correspondence:
\texttt{sean@mnemox.ai}. The author thanks Yijia Xiao (UCLA) for the arXiv
endorsement. This is a pre-registered, observational study on the full Hyperliquid
leaderboard universe: the cohort definition, detector, operating point, and claim
gates were frozen in a timestamped commit \emph{before} the sealed test split was
read once (Appendix~\ref{app:op}).}\\
Mnemox AI}

\date{\today}

\begin{document}
\maketitle

\begin{abstract}
A common, rarely-tested assumption behind copy-trading risk controls is that
``behavioral monitoring catches blow-ups'': that a master trader's conduct drifts
\emph{before} the equity damage that wipes out followers. We test this on real
on-chain money over the \emph{full} public Hyperliquid leaderboard: 37{,}895
addresses, filtered to 10{,}395 ``master-worth-copying'' accounts, yielding
\textbf{2{,}621 genuine loss-driven idiosyncratic blow-ups} across 48 independent
crash-days. On a model-free panel of 2{,}477 blow-up masters, behavioral drift
leads equity for only \textbf{9\%} (216 masters); \textbf{69\% blow up with no
gradual behavioral signal at all}, and the overall median lead is \emph{negative}
($-8$ days): when conduct moves, it typically \emph{lags} the equity rather than
leading it. A purpose-built three-axis mixture-SPRT detector, pre-registered with
its operating point frozen on a disjoint tuning split and read once against a
sealed test set, confirms the boundary rather than beating it: it achieves
\emph{chance-level} discrimination (AUC $0.49$) at $2\%$ recall. It does post the
lowest false-positive rate of any method ($0.040$, $\approx\!1.9\times$ below a
leverage tripwire), but at negligible recall that precision is of little practical
use. Behavioral early warning is therefore \emph{not} a viable general safety net
for copy-trading blow-ups: most are sudden, a plain equity / drawdown stop fires
as early on them, and the residual behavioral signal (gradual over-leveraging)
precedes fewer than one blow-up in ten.
\end{abstract}


\medskip
\noindent\textbf{JEL / arXiv:} q-fin.TR (Trading and Market Microstructure);
G11, G23, G41.

\section{Introduction}
\label{sec:intro}

Copy trading and social trading let retail followers mirror a ``master'' trader's
positions automatically. The arrangement industrializes a single account's risk:
when a master blows up, every follower blows up with them, on the same candle.
Regulators have repeatedly flagged this contagion channel and the asymmetry of
information it creates: a follower sees a headline return and a leaderboard rank,
not the master's moment-to-moment risk-taking
\citep{iosco2025imitative,esma2023copytrading,pelster2019socialtrading}.
The implicit promise of every ``risk score'' and ``copy-stop'' product layered on
top is that \emph{watching how the master trades} gives an earlier, cleaner warning
than watching the equity curve alone.

That promise rests on a load-bearing empirical assumption that is almost never
tested adversarially on real adversarial data: \textbf{behavioral drift precedes
the equity damage.} If a master's conduct degrades gradually---creeping leverage,
abandoned stops, averaging into losers---before the account craters, then a
behavioral monitor buys followers lead-time the equity itself cannot. If instead
most blow-ups are \emph{sudden}---one oversized trade, a gap, or a master who was
always reckless and simply got unlucky---then there is no behavioral runway to
detect, and a plain equity threshold is both simpler and earlier.

We put this assumption on real on-chain money. Hyperliquid is a high-volume
decentralized perpetual-futures venue whose public \texttt{info} API exposes, with
no authentication, the recent fill, order, and ledger history for any address
(subject to an $\approx$10{,}000-fill cap we return to in Section~\ref{sec:data}),
plus a coarse account-value history \citep{hyperliquid2024docs}. This is an unusually
transparent setting: positions and liquidations are settled on-chain, the
leaderboard is public, and there is no survivorship curation by a broker. We
freeze a leaderboard universe at a fixed date, label blow-ups by a forward-only
equity rule, and ask---first model-free, then with a purpose-built detector---when,
and for whom, behavior actually leads equity.

\paragraph{Contributions.}
\begin{enumerate}[leftmargin=1.4em,itemsep=2pt]
  \item \textbf{The boundary (the headline).} On 2{,}477 real idiosyncratic
        blow-up masters, behavioral drift leads equity for only $\approx$9\%; 69\%
        are sudden; the overall median lead is \emph{negative}. The popular
        ``behavioral monitoring catches blow-ups'' assumption is false as a general
        law, including in an earlier, scaled-cohort version of this project, whose
        more favorable numbers this full-universe run overturns.
  \item \textbf{A pre-registered detector that confirms it.} A three-axis
        mixture-SPRT detector, frozen on a disjoint tuning split and read once
        against a sealed test set, achieves \emph{chance-level} discrimination
        (AUC $0.49$) at $2\%$ recall: the behavioral machinery cannot rank a
        soon-to-blow-up master above a stable one. It does attain the lowest
        false-positive rate of any method ($\approx 1.9\times$ below a leverage
        tripwire), a real but, at this recall, inconsequential property.
  \item \textbf{What little survives is exposure.} Among the few masters that do
        drift early, the trigger is almost always the exposure (leverage) axis, not
        discipline or tilt---but it precedes fewer than one blow-up in ten, and our
        on-chain leverage proxy is coarse, so we do not over-read the mechanism.
\end{enumerate}

We are explicit about epistemic status. This is a \emph{pre-registered},
observational study on the \emph{full} leaderboard universe: the cohort definition,
detector form, operating point, and claim gates were frozen in a timestamped commit
\emph{before} the sealed test split was read (Appendix~\ref{app:op}). It overturns
an earlier exploratory, scaled-cohort version that had reported a usable
high-precision detector; the contamination that inflated that result---withdrawal
-driven equity craters mislabeled as trading blow-ups---is removed here by a
loss-confirmed labeling rule (Section~\ref{sec:data}). Honesty about where
behavioral early warning works \emph{and}, mostly, where it does not is the point of
the paper, not a caveat to it.

\section{Related work}
\label{sec:related}

\paragraph{Copy / social trading and its risks.}
Social-trading platforms have been studied mostly from the follower's behavioral
side---herding, the disposition effect amplified by visible reputation, and the
hazards of mirroring leveraged strategies
\citep{pelster2019socialtrading,barber2000trading}. Supervisory bodies have issued
explicit warnings about copy trading and ``finfluencer''-driven retail
participation, stressing the information asymmetry between the copied master and
the copier \citep{iosco2025imitative,esma2023copytrading}. Commercial copy
desks operationalize this with proprietary ``risk scores'' and automatic
copy-stops; a representative class is the forex copy-trading risk guards that monitor
a master and can pause copying. Such systems are typically closed, forex-centric,
and, critically, unvalidated in public against the alternative of simply watching
equity. Our detector targets the same job (drift / behavioral risk intelligence for
a broker or copy desk) but is vendor-neutral, on-chain, and reported with its
false-positive rate alongside its lead-time.

\paragraph{Changepoint and statistical process control.}
Detecting a shift in a stream is classical. The cumulative sum (CUSUM) chart and
its SPC relatives detect a step change in a monitored statistic
\citep{page1954cusum,montgomery2009spc}; Bayesian Online Changepoint Detection
(BOCPD) maintains a run-length posterior \citep{adams2007bocpd}. Our per-axis test
is a sequential probability ratio test \citep{wald1945sprt} in its mixture form
(mSPRT), which is ``always valid''---it controls error under continuous monitoring
and optional stopping, exactly the peeking problem that arises when an alarm is
checked every bucket \citep{johari2017peeking}. We compose three axes with a
Holm correction \citep{holm1979simple} and shrink each axis's self-baseline toward
a universe prior in James--Stein fashion \citep{james1961stein}.

\paragraph{Behavioral / concept drift.}
``Drift'' in a trader is a special case of concept drift---the statistical
properties of a stream changing over time \citep{gama2014concept}. The
machine-learning literature emphasizes detecting and adapting to drift; we instead
ask the prior question of whether a financially-meaningful drift even exists ahead
of the outcome it is supposed to predict.

\paragraph{Equity thresholds vs.\ behavioral detectors: reframed.}
In earlier internal work on self-monitoring trading agents we found that a simple
maximum-drawdown stop on the equity curve outperformed a CUSUM behavioral detector
for cutting losses, consistent with the treatment of drawdown as a primary risk
measure \citep{magdon2004maxdd}. We previously read this as a defeat for behavioral
methods. The present study reframes it correctly: \emph{equity-threshold methods
win in general because most blow-ups are sudden} (Section~\ref{sec:results}). When
the damage arrives without a gradual behavioral antecedent, nothing beats reading
the damage directly. Behavioral detectors could only add value within the gradual
subclass, and that subclass turns out to be both small (under one blow-up in ten)
and, even with a pre-registered detector, not separable from stable masters above
chance (Section~\ref{sec:results}). The reframing sharpens the negative rather than
rescuing the method: equity wins not by a narrow margin but because the behavioral
runway a detector would need is usually not there.

\paragraph{Methodological hygiene.}
Performance studies in finance are notoriously fragile to survivorship bias
\citep{brown1991survivorship}, multiple testing and backtest overfitting
\citep{harvey2015backtesting}, and clustered dependence. We address these with a
frozen-at-$\Tnull$ universe (no survival filter), a strict tuning/test split with
the operating point chosen on the tuning split only---before the sealed test data
were read---and event-clustered bootstrap
inference \citep{efron1986bootstrap,cameron2008clusterboot} that resamples
crash-days rather than masters.

\section{Data}
\label{sec:data}

\paragraph{Source.}
All data come from the public Hyperliquid \texttt{info} API
\citep{hyperliquid2024docs}, which requires no authentication: per-address fills
(\texttt{userFills}, carrying a \texttt{liquidation} flag), orders
(\texttt{historicalOrders}, including trigger / reduce-only metadata), ledger
updates (\texttt{userNonFundingLedgerUpdates}, including margin top-ups and
withdrawals), and a coarse account-value history (\texttt{portfolio}). Requests are
throttled ($\geq 0.4$\,s spacing) with retry on rate-limit, and raw responses are
archived for reproducibility (Appendix~\ref{app:repro}).

\paragraph{Frozen universe.}
To kill survivorship bias we freeze the universe at $\Tnull = $ 2025-06-01 UTC and
look forward to a window end of 2026-04-30 UTC ($\approx$11 months). The universe
is a leaderboard snapshot filtered to addresses with a pre-$\Tnull$ peak perpetual
account value $\geq \$25{,}000$ (a ``master worth copying'' floor that removes
dust) and at least two pre-$\Tnull$ equity points. Selection uses \emph{only}
information available at $\Tnull$; no outcome (survival) filter is applied. We
process the \emph{entire} leaderboard (37{,}895 addresses) and retain the
\textbf{10{,}395} that clear the floor. Crucially, the filter is on the
\emph{pre-}$\Tnull$ \emph{peak}, not on the current account value: masters who have
since blown up below \$25k are kept rather than silently dropped, which is the
selection error that would otherwise hide exactly the events we study.

\paragraph{Labeling (forward-only, loss-confirmed).}
A master is labeled a blow-up by a forward-only, \emph{loss-confirmed} rule
(Appendix~\ref{app:rule}): scanning forward from $\Tnull$, the first peak-to-trough
drawdown exceeding 70\% with no recovery to 80\% of the prior peak within 30 days
\emph{whose cumulative-PnL drop is at least half the equity drop}. The
loss-confirmation is not cosmetic. At $\Tnull = $ 2025-06-01, \textbf{65\% of naive
equity-only craters are withdrawals, not losses} (median PnL-to-equity-drop ratio
$0.21$): the account value falls because the master moved capital out, not because
trading lost it. A withdrawal crater therefore resets the running peak and the scan
continues, so the labeled time is the first genuinely \emph{loss-driven}
account-death, whenever it occurs. (An earlier scoping pass at a later $\Tnull$ had
found most craters already loss-confirmed; that does not hold at our frozen date,
which is why the filter is enforced \emph{inside} the labeler rather than as an
afterthought.) Among the retained blow-ups the median loss-to-equity ratio is
$1.0$. The labeling looks forward only to \emph{confirm} a crater (the 30-day
recovery and loss-confirmation checks); the labeled trigger time itself is the crater
onset, and because the detector is evaluated only strictly \emph{before} that time
under the guard band (Section~\ref{sec:method}), no label information leaks into the
detection window.

\paragraph{Idiosyncratic vs.\ market-event blow-ups.}
A market-wide dump synchronizes many masters on the same UTC day; such days are not
``predict the blow-up'' events but common-shock events. We tag a UTC day as a
\emph{market-event} day when at least $3\%$ of the \emph{active} universe---masters
with an equity observation in the surrounding window---craters on that day, and
separate those blow-ups from idiosyncratic ones. This yields \textbf{2{,}621
idiosyncratic blow-ups across 48 independent crash-days}, with the market-event
days (2025-06-04, 2025-06-11, 2025-10-15) tagged and held out of the idiosyncratic
analysis. The effective sample size for inference is the number of independent
crash-day clusters, not the raw blow-up count---hence the event-clustered
statistics in Section~\ref{sec:method}.

\paragraph{Behavioral sub-cohort.}
Equity labeling needs one cheap call per address; the full behavioral trajectory
(fills, orders, ledger) is expensive, so we pull it only for a labeled sub-cohort.
This \emph{behavioral sub-cohort} comprises \textbf{3{,}455 masters (2{,}482
blow-up / 973 stable)}: the idiosyncratic blow-ups whose complete behavioral
trajectory could be fetched within the rate-limited collection window (2{,}482 of
the 2{,}621 labeled; the remainder timed out against the fill-history cap and the
API throttle), plus a seed-fixed (address-hashed) sample of stable masters; the
resulting cohort is blow-up-heavy (2{,}482 blow-up / 973 stable, not a balanced
design). We do not balance the classes: FPR is reported on stable masters and recall
on blow-ups separately, with a rank-based AUC that is insensitive to the class ratio.
An address-hashed split assigns 1{,}405 masters to tuning, 656 to a validation split
used only for pre-lock development checks (it reports no numbers here), and a
\emph{sealed} test set of 1{,}394 (990 blow-up / 404 stable). The operating point is chosen on the tuning split
alone; all Section~\ref{sec:results} numbers are computed on the sealed test set,
read exactly once after the detector was frozen (Appendix~\ref{app:op}).

\paragraph{The 10k-fill-cap selection line.}
Hyperliquid's fill history is capped at the most recent $\approx$10{,}000 fills per
address. For very high-frequency masters this window does not reach back to
$\Tnull$, so a pre-$\Tnull$ behavioral baseline cannot be built from fills. Rather
than drop these masters (which would bias toward low-frequency traders---a
disclosed selection effect), we estimate each master's baseline from its
\emph{earliest in-observation healthy segment} (Section~\ref{sec:method}). We name
this line explicitly because it biases the baseline toward the in-window early
period and we report the fraction affected.

\paragraph{Ethics and pseudonymization.}
All addresses are public pseudonymous on-chain identifiers; no personal identity is
collected, inferred, or published. We report only aggregate statistics and present
no address-level case study. No trading is performed and no capital is at
risk; the study is purely observational on already-public ledger data.

\section{Method}
\label{sec:method}

\subsection{Three orthogonal behavioral axes}
We summarize a master's order flow on three volatility-normalized axes, each chosen
so that a deteriorating master moves it in a known sign:
\begin{description}[leftmargin=1.6em,itemsep=2pt]
  \item[Exposure ($\uparrow$ is worse).] Leverage, notional growth, and trade
        size-in-sigma (size scaled by the realized volatility of the traded
        instrument). Captures gradual over-leveraging.
  \item[Discipline ($\downarrow$ is worse).] Stop-attach rate, reduce-only rate,
        and holding time. Captures the abandonment of risk controls.
  \item[Tilt ($\uparrow$ is worse).] Margin top-ups, adding-into-losers, and
        trade-rate spikes. Captures emotional / revenge trading.
\end{description}
Each raw primitive is computed per time bucket (4-hour buckets in the reported
configuration) from the trajectory.

\subsection{Early-window self-baseline with shrinkage}
Because the fill cap denies many masters a pre-$\Tnull$ baseline
(Section~\ref{sec:data}), each axis is $z$-scored against an \emph{early-window
self-baseline}: the master's earliest contiguous \emph{healthy} prefix---the run of
buckets with equity $\geq 80\%$ of the running peak, ending at the first sustained
drop and capped at half the series. Both the running peak and the drop are computed
online from the prefix, so the baseline uses no future information. Masters with too
short a healthy prefix fall back to a short best-effort early window rather than being
excluded; we report this stability accounting below. To stabilize small-sample
baselines we shrink each master's per-axis location toward a tuning-split universe
distribution by James--Stein shrinkage \citep{james1961stein} with intensity $\kappa$.

\subsection{Per-axis mixture-SPRT and composite alarm}
On each axis we run a one-sided mixture sequential probability ratio test
\citep{wald1945sprt,johari2017peeking} in the deterioration direction, with
per-axis level $\alpha = 0.01$ and a per-axis mixing scale $\tau$---the scale of the
mSPRT's mixing distribution over the drift effect size ($\tau = 0.3$ in the reported
configuration). The mSPRT is ``always valid'': checking it every
bucket does not inflate the error rate, which is essential because an alarm is, by
construction, monitored continuously \citep{johari2017peeking}. The composite alarm
is a Holm-corrected \citep{holm1979simple} min-gate across the three axes,
\emph{sustained} for $M$ consecutive buckets:
\begin{equation}
\text{alarm}_t \;=\; \mathbf{1}\!\left[
\text{Holm}\!\big(p^{\text{exp}}_t, p^{\text{disc}}_t, p^{\text{tilt}}_t\big)
\text{ rejects for } M \text{ consecutive buckets}\right].
\end{equation}

\subsection{Guard band (anti-label-leakage)}
Exposure and tilt are mechanically coupled to equity, so an ``early'' alarm fired
\emph{after} the damage has begun would be a self-fulfilling label leak. We
therefore impose a guard band: an alarm counts as a genuine early warning only
while equity $\geq 70\%$ of the running peak \emph{and} strictly before the first
liquidation fill. Because alarms are scored only \emph{after} the burn-in baseline
window, the baseline-estimation segment (the earliest healthy buckets) and the
counted-alert region (after burn-in, up to 70\%-of-peak / first liquidation) do not
overlap in time.

\subsection{Baselines}
We compare against three non-trivial early tripwires, all run under the
\emph{same} guard band so the comparison is apples-to-apples:
\begin{description}[leftmargin=1.6em,itemsep=2pt]
  \item[B1 --- leverage-p90 tripwire.] Fire when leverage exceeds the master's own
        90th percentile.
  \item[B2 --- drawdown-velocity.] Fire on accelerating drawdown velocity.
  \item[B3 --- leverage-up-and-add.] Fire when leverage rises while the master adds
        into a losing position.
\end{description}
The central methodological point: \textbf{comparing lead-time without also
reporting FPR is meaningless when a baseline near-always-fires.} A tripwire that
alerts on essentially every master will trivially ``lead'' every blow-up by a long
margin while being useless in deployment. We therefore report, for every method,
the full triple $(\fpr,\ \text{recall},\ \text{lead})$ on the stable and blow-up
cohorts jointly.

\subsection{Operating-point selection (no test snooping)}
We grid-search only two knobs on the tuning split---sustain $M \in \{2,3,4\}$ and
shrinkage $\kappa \in \{7,14\}$---holding the mixing scale $\tau=0.3$, 4-hour buckets,
and a 20-bucket burn-in fixed a priori. Among grid points with tuning-split stable-FPR
$\leq 0.10$ we take the highest-recall configuration, breaking ties by the longest
tuning lead. This selects $M=3$, $\kappa=7$ (with the fixed $\tau$, bucket size, and
burn-in)
(Appendix~\ref{app:op}), which---together with the detector form and the claim
gates---was frozen in a timestamped commit \emph{before} the sealed test split was
read. Every number in Section~\ref{sec:results} is then computed on that sealed
test set, read exactly once. We note that $\kappa$ is effectively inert here
because the self-baseline and the universe prior nearly coincide on-chain
(Section~\ref{sec:discussion}); its frozen value carries no signal.

\subsection{Event-clustered inference}
Blow-ups cluster in time: a handful of crash-days produce many of them, so masters
are not independent observations. All confidence statements resample the
\emph{crash-day clusters} (not the masters) with 5{,}000 bootstrap replicates
\citep{efron1986bootstrap,cameron2008clusterboot}; in the blow-up-versus-stable AUC,
blow-ups are resampled by crash-day cluster and each stable master as its own
singleton cluster. We report the effective number of independent clusters as the true
power unit. Leave-one-event-out
robustness is computed analogously.

\section{Results}
\label{sec:results}

\subsection{Claim 1 --- the boundary: behavior does not generally lead equity}
\label{sec:boundary}

We first test the load-bearing assumption \emph{model-free}, with no detector. For
each of 2{,}477 real idiosyncratic blow-up masters (those with $\geq$20 behavioral
buckets; 5 of the 2{,}482 fetched blow-ups fall short and drop from this panel) we
mark $T_{\text{behavior}}$, the first bucket after the early-window
baseline at which behavior clearly degrades---operationally, leverage exceeding
$1.5\times$ the baseline's 90th-percentile leverage, \emph{or} leverage above that
90th percentile while the master adds to a losing position, \emph{or} the stop-attach
rate falling below the baseline's 10th percentile---and $T_{\text{equity}}$, the first
bucket where equity falls below 90\% of its running peak. The lead is
$T_{\text{equity}} - T_{\text{behavior}}$ (positive when behavior leads); a master
with no such degradation has no $T_{\text{behavior}}$ and counts as \emph{sudden}.

\begin{figure}[t]
  \centering
  \includegraphics[width=0.82\linewidth]{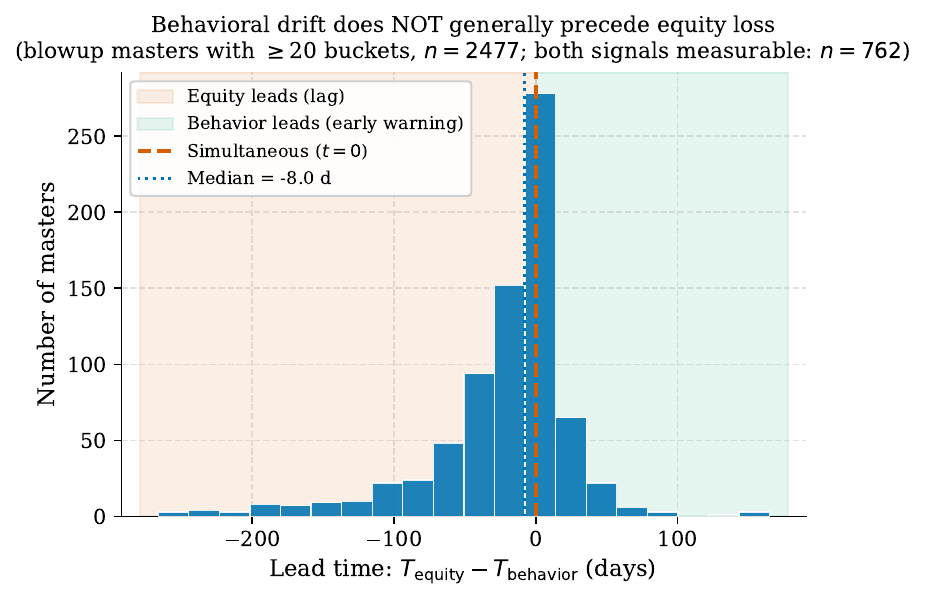}
  \caption{\textbf{The boundary (Claim 1).} Distribution of the behavior-minus-equity
  lead over the 762 of 2{,}477 idiosyncratic blow-up masters with both a
  behavioral-degradation signal and a measurable equity crossing (model-free, no
  detector). A further 1{,}706 (69\%) show no behavioral signal at all---the
  \emph{sudden} blow-ups, which cannot be placed on this axis---and 9 have a signal but
  no equity crossing within the fill-capped behavioral window and are omitted
  ($762+1{,}706+9 = 2{,}477$). The mass at
  and below zero shows that behavioral drift does \emph{not} generally precede equity
  damage: 216 of all 2{,}477 (9\%) have a positive lead, and the median over the
  measurable cases is \emph{negative} ($\approx-8$\,d). The thin right tail is the
  gradual-leverage minority ($\approx$13-day median lead).}
  \label{fig:boundary}
\end{figure}

The result is the paper's headline negative (Figure~\ref{fig:boundary}). Behavior
leads equity for only \textbf{216 of 2{,}477 ($\approx$9\%)} masters. Fully
\textbf{69\% blow up with no gradual behavioral signal at all}---they are sudden:
one outsized trade, a gap, or a master who was aggressive throughout with no
``drift'' to detect. Over the 762 cases where both timestamps exist, the median lead
is \emph{negative} (\textbf{$-8$ days}): when behavior moves at all, it more often
\emph{lags} the equity than leads it. Only in the minority right tail is there real
runway---there the median lead is $\approx$\textbf{13 days} (IQR 7.3--24.5\,d), and
it is almost entirely an exposure phenomenon (\textbf{215 of 216} are carried by the
leverage axis, not discipline or tilt).

The boundary is clear: \emph{behavioral monitoring is not a general
early-warning solution for copy-trading blow-ups.} A behavioral detector that
assumes gradual, sustained drift cannot help with the majority of blow-ups, and a
plain equity threshold will fire as early or earlier on the sudden ones. This is
the boundary; the rest of the paper lives inside it.

\subsection{Claim 2 --- the pre-registered detector confirms the boundary}
\label{sec:fair}

\begin{figure}[t]
  \centering
  \includegraphics[width=0.72\linewidth]{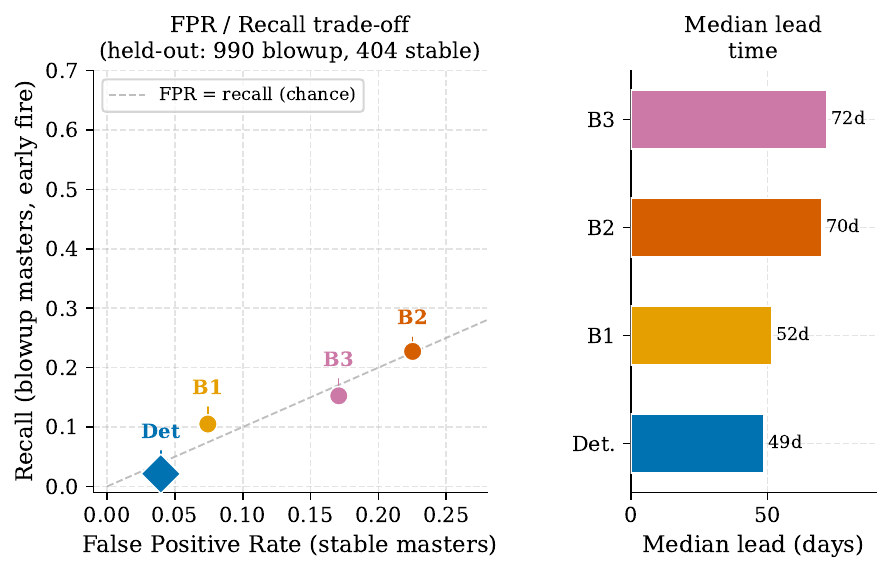}
  \caption{\textbf{Fair comparison in FPR--recall space (Claim 2).} Each method is a
  point; the top-left corner (low FPR, high recall) is best. The detector sits at the far left (FPR 0.040) but
  near the floor (recall 0.021): the lowest false-alarm rate, almost no coverage,
  and AUC at chance. B1 buys $\approx 5\times$ the recall at $\approx 1.9\times$ the
  FPR; B2/B3 buy more recall only by alerting on 17--23\% of stable masters. Reading
  lead-time without the FPR axis (Table~\ref{tab:fair}) would reward the
  fire-on-everything baselines for an artifact.}
  \label{fig:fpr}
\end{figure}

A purpose-built behavioral detector might still beat the crude model-free rule. It
does not. On the sealed test set (990 blow-up / 404 stable), with every method under
the same guard band and the detector at its frozen operating point ($M=3$,
$\kappa=7$, $\tau=0.3$, 4-hour buckets), the three-axis mixture-SPRT achieves
\emph{chance-level} discrimination: its continuous drift-evidence score separates
blow-up from stable masters with $\mathrm{AUC}=0.49$ (event-clustered 95\% CI
$[0.47,\,0.51]$, straddling the $0.5$ chance line), and as a binary alarm it fires
early on only $2\%$ of blow-ups. The
full $(\fpr,\ \text{recall},\ \text{lead})$ triple for each method is in
Table~\ref{tab:fair} and Figure~\ref{fig:fpr}.

\begin{table}[t]
  \centering
  \caption{\textbf{Fair held-out comparison (Claim 2).} 990 blow-up / 404 stable,
  the \emph{sealed} test set; operating point frozen on a disjoint tuning split
  before the test was read. FPR is measured on stable masters; recall is the
  fraction of blow-ups with a guard-banded early alarm; lead is the median
  alarm-to-crater time among caught blow-ups. The detector posts the lowest FPR
  ($\approx 1.9\times$ below B1) but at only $2\%$ recall, and its $\mathrm{AUC}=0.49$
  is at chance: low FPR with negligible recall is not a usable detector. The
  baselines' longer ``leads'' are a fire-on-everything artifact---they alert on
  17--23\% of \emph{stable} masters.}
  \label{tab:fair}
  \small
  \setlength{\tabcolsep}{5pt}
  \begin{tabular}{@{}lccccl@{}}
    \toprule
    Method & FPR $\downarrow$ & Recall $\uparrow$ & Median lead & FP / 404 & Note \\
    \midrule
    \textbf{Detector (mSPRT, 3-axis)} & \textbf{0.040} & 0.021 & 49 d & 16 & lowest FPR; chance AUC \\
    B1 \;leverage-p90 tripwire        & 0.074 & 0.105 & 52 d & 30 & $1.9\times$ FPR, $5\times$ recall \\
    B2 \;drawdown-velocity            & 0.225 & 0.227 & 70 d & 91 & fires widely \\
    B3 \;leverage-up-and-add          & 0.171 & 0.153 & 72 d & 69 & fires widely \\
    \bottomrule
  \end{tabular}
\end{table}

Three readings follow.
\begin{enumerate}[leftmargin=1.4em,itemsep=2pt]
  \item \textbf{Lowest FPR, but it buys nothing.} The detector false-alarms on 16 of
        404 stable masters ($\fpr = 0.040$) versus 30--91 for the baselines
        ($\fpr = 0.074$--$0.225$)---the lowest by $\approx 1.9\times$. In isolation
        that precision is attractive; paired with $2\%$ recall it means the detector
        is silent on essentially every blow-up that matters.
  \item \textbf{Chance-level discrimination.} The continuous drift-evidence score
        ranks blow-up above stable masters no better than a coin
        ($\mathrm{AUC}=0.49$, 95\% CI $[0.47,\,0.51]$). The behavioral
        signal does not separate the two populations---the pre-registered
        confirmation of Claim 1.
  \item \textbf{The baselines' ``leads'' are an artifact.} B2 and B3 post longer
        median leads (70 d, 72 d) and higher recall only by firing on 17--23\% of
        \emph{stable} masters---near-always-firing. Lead-time read without FPR rewards
        exactly this pathology; with FPR shown, the artifact is plain.
\end{enumerate}

Taken together, the detector does \textbf{not} discriminate: its $2\%$ recall and
chance-level AUC leave it silent on essentially all blow-ups, and its one favorable
property---a low false-positive rate---is of little operational value at this
coverage. This is the Claim-1 boundary, confirmed by the very machinery built to
beat it.

\subsection{Lead-time, where it exists}
\label{sec:leadtime}

\begin{figure}[t]
  \centering
  \includegraphics[width=0.72\linewidth]{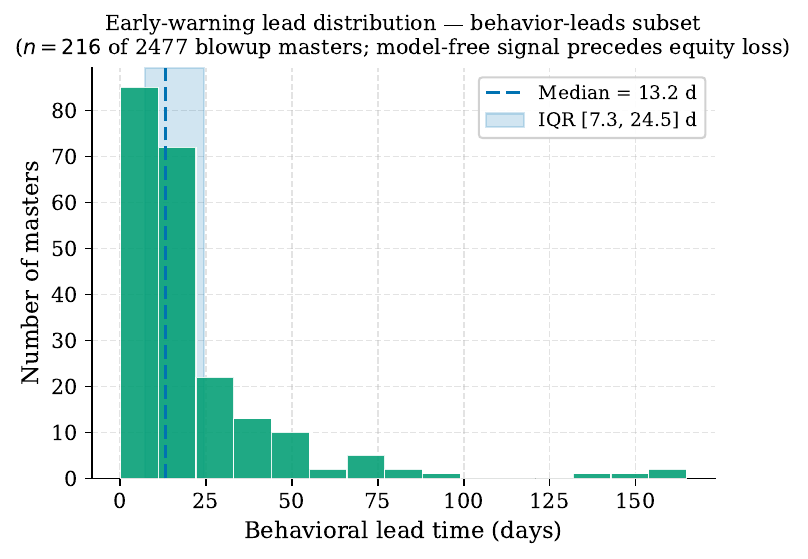}
  \caption{\textbf{Lead-time, where it exists (Claim 2).} Model-free
  behavior-minus-equity lead over the 216 idiosyncratic blow-up masters with a
  \emph{positive} lead. Median $\approx$13 days, right-skewed. Long where it
  exists---but conditioned on the $\approx$9\% of blow-ups that show any gradual
  signal; it is silent on the rest. The signal here is the model-free drift rule
  (leverage / loser-add / stop-attach versus an early-window threshold), \emph{not}
  the mSPRT detector, whose median lead on its own fired subset is $\approx$49 days.}
  \label{fig:lead}
\end{figure}

Among the minority of blow-ups that \emph{do} show an early behavioral signal, the
lead can be long. Figure~\ref{fig:lead} plots the model-free behavior-minus-equity
lead over the 216 masters with a positive value: a median of $\approx$13 days
(IQR 7.3--24.5\,d), right-skewed, some masters slowly over-leveraging for months
before the crater. This is the most that can be said for behavioral early warning,
and it must be read with its denominator---it is conditioned on the $\approx$9\% of
blow-ups that drift at all. The detector, on the $2\%$ it fires on, shows a
similarly long median ($\approx$49 days); but conditioning on $2\%$ coverage makes
that number a curiosity, not a value proposition.

\subsection{Claim 3 --- what little survives is exposure}
\label{sec:mechanism}

Which axis carries what signal there is? Almost entirely \textbf{exposure}: in the
model-free boundary analysis, 215 of the 216 behavior-leads masters are triggered by
the leverage axis rather than by discipline (stop-attach, reduce-only) or tilt
(margin top-ups, loser-adding). Two caveats keep this from being a clean
``mechanism.'' First, it describes only the $\approx$9\% gradual minority, not
blow-ups at large. Second, our on-chain leverage signal is a coarse,
volatility-normalized order-flow proxy, not reconstructed account leverage, and the
discipline and tilt primitives are sparse and noisy in public fills, so the
dominance of exposure may partly reflect what is cleanly \emph{measurable} on-chain
rather than a behavioral truth. We therefore report exposure as the surviving
correlate, not as a validated causal channel, and omit a single-master case study
that the noisy leverage proxy cannot support responsibly.

\section{Discussion}
\label{sec:discussion}

\paragraph{Why equity beats behavior.}
Equity thresholds win (Section~\ref{sec:related}) \emph{because} most blow-ups are
sudden (Claim 1) and because, even among the gradual minority, a detector
pre-registered and read on a sealed test cannot rank them above stable masters at
better than chance (Claim 2). Our earlier internal reading---``drawdown stops beat a
behavioral CUSUM, so behavioral methods lose''---turns out to have been right, for
the right reason, on a far larger and cleaner sample: when the damage arrives with no
gradual antecedent, nothing beats reading the damage directly, and the antecedent is
usually absent. The conditional that survives is narrow---watch the equity; a
behavioral layer earns its place only if one can first define, causally and ahead of
time, the small subclass that drifts.

\paragraph{Implications for copy-desk risk controls.}
The practical message is deflationary. A behavioral overlay that fires on $2\%$ of
blow-ups and ranks them at chance does not earn its place as a general early-warning
layer, however low its false-positive rate. A broker or copy desk is better served
by a plain equity / drawdown stop, which catches the sudden majority directly and at
least as early. The $4\%$ false-positive rate we measure is real and would matter
\emph{if} it came with usable recall; it does not, so it does not rescue the
approach. Behavioral monitoring is, at most, a narrow complement for a
gradual-leverage subclass that is too small, and too hard to define ahead of time,
to anchor a product.

\paragraph{Honest limitations.}
We foreground these because they bound every claim above.
\begin{itemize}[leftmargin=1.4em,itemsep=2pt]
  \item \textbf{Near-zero recall.} The detector fires early on $\approx$2\% of
        blow-ups and discriminates at chance (AUC $0.49$). It is not a general
        copy-trading safety net and must not be sold as one.
  \item \textbf{Coarse on-chain proxies.} The leverage signal is a
        volatility-normalized order-flow statistic, not reconstructed account
        leverage; discipline (stop-attach, reduce-only) and tilt (top-ups,
        loser-adding) rely on sparse, partly opaque public-fill metadata. The
        dominance of the exposure axis may partly reflect what is cleanly measurable
        rather than a behavioral truth.
  \item \textbf{Labeling choices.} Blow-ups are equity-drawdown events with a
        loss-confirmation filter; the $70\%$ threshold, the $0.5$ PnL-to-equity ratio,
        and the early-window self-baseline are defensible but not unique. Robustness
        reruns over the drawdown threshold are noted in Appendix~\ref{app:rule}.
  \item \textbf{Post-hoc subclass.} The ``gradual-leverage'' subclass is identified
        from observed drift and then shown to be (weakly) detectable by a drift
        detector---a circularity. A \emph{causal, pre-registered} definition of the
        subclass, fixed before the outcome is read, is the central open problem.
  \item \textbf{$\kappa$ inert.} Because on-chain masters' self-baselines and the
        universe prior nearly coincide, the shrinkage intensity $\kappa$ does almost
        nothing here; its frozen value carries no signal.
  \item \textbf{Single venue, single window.} Hyperliquid, $\Tnull = $ 2025-06-01 to
        2026-04-30. Generalization to other venues, asset classes, and regimes is
        untested.
\end{itemize}

\paragraph{Future work.}
The decisive open problem is a \emph{causal, pre-registered} definition of the
gradual-leverage subclass---a leverage-trajectory criterion fixed before outcomes
are read---so that ``a detector catches the gradual subclass'' becomes a falsifiable
prediction rather than a tautology, and so the subclass can be flagged in real time
rather than in hindsight. Reconstructing true account leverage from positions and
margin (rather than order-flow proxies) would sharpen the exposure axis; widening the
window and replicating across venues and asset classes would test whether the
boundary we report is a property of this market or of copy-trading blow-ups in
general.

\section{Conclusion}
\label{sec:conclusion}

We tested, on real on-chain money and at full-universe scale, the assumption that
behavioral monitoring catches copy-trading blow-ups. It does not. On 2{,}477
idiosyncratic blow-up masters, behavioral drift leads equity for only $\approx$9\%
(216), 69\% are sudden, and the median lead is negative. A three-axis mixture-SPRT
detector, pre-registered and read once against a sealed test set, confirms it:
chance-level discrimination (AUC $0.49$) at $2\%$ recall, its lone virtue a
lowest-in-class false-positive rate that negligible coverage renders moot. The
residual signal, where it exists, is gradual over-leveraging---but it precedes fewer
than one blow-up in ten and rests on a coarse proxy. The result is a clear boundary:
behavioral early warning is not a general safety net for
copy-trading blow-ups; most are sudden, an equity stop fires directly when the damage
starts, and a behavioral layer would first need a causal, ahead-of-time definition of
the small drifting
subclass before it could earn its place. The result overturns a smaller, exploratory
version of this project---a reminder that survivorship- and withdrawal-contaminated
cohorts can manufacture a signal that a clean, full-universe test dissolves.

\appendix
\section{Reproducibility}
\label{app:repro}

All inputs are public and unauthenticated. We use the Hyperliquid \texttt{info}
endpoint \citep{hyperliquid2024docs} for fills, orders, ledger updates, and the
\texttt{portfolio} account-value history, and the public leaderboard snapshot for
the universe. The fetcher throttles to $\geq 0.4$\,s between calls with retry on
HTTP-429, and raw JSON responses are archived so the cohort can be rebuilt without
re-querying. The detector (per-trade primitives $\to$ bucket-close per-axis
observations $\to$ per-axis mSPRT $\to$ Holm min-gate, sustained-$M$ composite $\to$
70\%-peak / pre-first-liquidation guard band), the baselines B1--B3, the early-window
baseline with stability guard, and the event-clustered bootstrap are released as
code alongside this paper. The cohort definition (universe filter, $\Tnull$, window,
labeling rule, market-event tag) and the detector form are pre-specified; the
operating point is chosen on the tuning split only and was frozen---with the
detector form and the claim gates---in a timestamped version-control commit
(\texttt{573110e}) before the sealed test split was read once (Appendix~\ref{app:op}).

\section{The forward-only blow-up rule}
\label{app:rule}

Scanning forward from $\Tnull$, a candidate blow-up is the first equity bar at which
the drawdown from the running pre-$t$ peak exceeds 70\% with no recovery to 80\% of
that peak within 30 days. A candidate is \emph{confirmed} only if its crater is
loss-driven: the cumulative-PnL drop from peak to trough must be at least half the
equity drop. A candidate that fails this test is a withdrawal, not a loss---the
running peak is reset to the post-withdrawal level and the scan continues---so the
labeled time is the first genuinely loss-driven account-death. The rule uses
information after $t$ only to \emph{confirm} the crater (the 30-day recovery and
loss-confirmation checks); because the detector is evaluated strictly before the
crater onset under the guard band, no label information leaks into the detection
window. Robustness reruns vary the drawdown threshold over $\{0.50, 0.70, 0.85\}$.

\section{Operating-point selection procedure}
\label{app:op}

The detector grid varies sustain $M \in \{2,3,4\}$ and shrinkage
$\kappa \in \{7,14\}$ at fixed $\tau = 0.3$, 4-hour buckets, and a 20-bucket
burn-in. On the \emph{tuning} split only (1{,}025 blow-up / 380 stable) we retain
grid points with tuning stable-FPR $\leq 0.10$, select the highest-recall
configuration, and break ties by the longest tuning lead. This yields the operating
point $M=3$, $\kappa=7$, which, with the detector form and the claim gates, was
frozen in a timestamped pre-registration commit before any test data were read. The
\emph{sealed} test set (990 blow-up / 404 stable) is then read exactly once with this
frozen tuple; the tuning re-derivation is verified to reproduce the frozen value, and
no re-tuning is performed on the test. Under this operating point the test blow-up
cohort splits into 822 with a clean early-window baseline and 168 falling back to the
relaxed baseline, none excluded outright---the relaxed stability guard converts a
previously high exclusion rate into a reported fallback rather than dropped masters.
As noted in Section~\ref{sec:method}, $\kappa$ is effectively inert here because the
self and universe distributions nearly coincide on-chain, so its frozen value carries
no signal.

\bibliographystyle{plainnat}
\bibliography{refs}

\begin{thebibliography}{18}
\providecommand{\natexlab}[1]{#1}
\providecommand{\url}[1]{\texttt{#1}}
\expandafter\ifx\csname urlstyle\endcsname\relax
  \providecommand{\doi}[1]{doi: #1}\else
  \providecommand{\doi}{doi: \begingroup \urlstyle{rm}\Url}\fi

\bibitem[Adams and MacKay(2007)]{adams2007bocpd}
Ryan~Prescott Adams and David J.~C. MacKay.
\newblock Bayesian online changepoint detection.
\newblock arXiv:0710.3742, 2007.

\bibitem[Barber and Odean(2000)]{barber2000trading}
Brad~M. Barber and Terrance Odean.
\newblock Trading is hazardous to your wealth: The common stock investment
  performance of individual investors.
\newblock \emph{The Journal of Finance}, 55\penalty0 (2):\penalty0 773--806,
  2000.

\bibitem[Brown et~al.(1992)Brown, Goetzmann, Ibbotson, and
  Ross]{brown1991survivorship}
Stephen~J. Brown, William Goetzmann, Roger~G. Ibbotson, and Stephen~A. Ross.
\newblock Survivorship bias in performance studies.
\newblock \emph{The Review of Financial Studies}, 5\penalty0 (4):\penalty0
  553--580, 1992.

\bibitem[Cameron et~al.(2008)Cameron, Gelbach, and
  Miller]{cameron2008clusterboot}
A.~Colin Cameron, Jonah~B. Gelbach, and Douglas~L. Miller.
\newblock Bootstrap-based improvements for inference with clustered errors.
\newblock \emph{The Review of Economics and Statistics}, 90\penalty0
  (3):\penalty0 414--427, 2008.

\bibitem[Efron and Tibshirani(1986)]{efron1986bootstrap}
Bradley Efron and Robert Tibshirani.
\newblock Bootstrap methods for standard errors, confidence intervals, and
  other measures of statistical accuracy.
\newblock \emph{Statistical Science}, 1\penalty0 (1):\penalty0 54--75, 1986.

\bibitem[{European Securities and Markets
  Authority}(2023)]{esma2023copytrading}
{European Securities and Markets Authority}.
\newblock Supervisory briefing on supervisory expectations in relation to firms
  offering copy trading services.
\newblock Supervisory Briefing ESMA35-42-1428, ESMA, Paris, 2023.

\bibitem[Gama et~al.(2014)Gama, {\v{Z}}liobait{\.e}, Bifet, Pechenizkiy, and
  Bouchachia]{gama2014concept}
Jo{\~a}o Gama, Indr{\.e} {\v{Z}}liobait{\.e}, Albert Bifet, Mykola Pechenizkiy,
  and Abdelhamid Bouchachia.
\newblock A survey on concept drift adaptation.
\newblock \emph{ACM Computing Surveys}, 46\penalty0 (4):\penalty0 1--37, 2014.

\bibitem[Harvey and Liu(2015)]{harvey2015backtesting}
Campbell~R. Harvey and Yan Liu.
\newblock Backtesting.
\newblock \emph{The Journal of Portfolio Management}, 42\penalty0 (1):\penalty0
  13--28, 2015.

\bibitem[Holm(1979)]{holm1979simple}
Sture Holm.
\newblock A simple sequentially rejective multiple test procedure.
\newblock \emph{Scandinavian Journal of Statistics}, 6\penalty0 (2):\penalty0
  65--70, 1979.

\bibitem[{Hyperliquid}(2024)]{hyperliquid2024docs}
{Hyperliquid}.
\newblock Hyperliquid documentation: Info endpoint and public {API}.
\newblock \url{https://hyperliquid.gitbook.io/hyperliquid-docs}, 2024.
\newblock Accessed 2026-05.

\bibitem[{International Organization of Securities
  Commissions}(2025)]{iosco2025imitative}
{International Organization of Securities Commissions}.
\newblock Online imitative trading practices: Copy trading, mirror trading,
  social trading.
\newblock Final Report FR/06/2025, IOSCO, Madrid, 2025.
\newblock Supersedes consultation report CR/10/2024.

\bibitem[James and Stein(1961)]{james1961stein}
W.~James and Charles Stein.
\newblock Estimation with quadratic loss.
\newblock \emph{Proceedings of the Fourth Berkeley Symposium on Mathematical
  Statistics and Probability}, 1:\penalty0 361--379, 1961.

\bibitem[Johari et~al.(2015)Johari, Pekelis, and Walsh]{johari2017peeking}
Ramesh Johari, Leo Pekelis, and David~J. Walsh.
\newblock Always valid inference: Bringing sequential analysis to {A/B}
  testing.
\newblock arXiv:1512.04922, 2015.
\newblock Mixture sequential probability ratio test (mSPRT) for continuous
  monitoring.

\bibitem[Magdon-Ismail and Atiya(2004)]{magdon2004maxdd}
Malik Magdon-Ismail and Amir~F. Atiya.
\newblock Maximum drawdown.
\newblock \emph{Risk Magazine}, 17\penalty0 (10):\penalty0 99--102, 2004.

\bibitem[Montgomery(2009)]{montgomery2009spc}
Douglas~C. Montgomery.
\newblock \emph{Introduction to Statistical Quality Control}.
\newblock John Wiley \& Sons, Hoboken, NJ, 6 edition, 2009.

\bibitem[Page(1954)]{page1954cusum}
E.~S. Page.
\newblock Continuous inspection schemes.
\newblock \emph{Biometrika}, 41\penalty0 (1/2):\penalty0 100--115, 1954.

\bibitem[Pelster and Hofmann(2018)]{pelster2019socialtrading}
Matthias Pelster and Annette Hofmann.
\newblock About the fear of reputational loss: Social trading and the
  disposition effect.
\newblock \emph{Journal of Banking \& Finance}, 94:\penalty0 75--88, 2018.

\bibitem[Wald(1945)]{wald1945sprt}
Abraham Wald.
\newblock Sequential tests of statistical hypotheses.
\newblock \emph{The Annals of Mathematical Statistics}, 16\penalty0
  (2):\penalty0 117--186, 1945.

\end{thebibliography}

\end{document}